# Characteristics and Origin of the Quadruple System at Pluto

S.A. Stern[1], H.A. Weaver[2], A.J. Steffl[1], M.J. Mutchler[3], W.J. Merline[1], M.W. Buie[4], E.F. Young[1], L.A. Young[1], & J.R. Spencer[1]

[1]Southwest Research Institute, 1050 Walnut St., Suite 400, Boulder, CO, 80302. [2]The Johns Hopkins Applied Physics Laboratory, 11100 Johns Hopkins Road, Laurel, MD 20723, [3]Space Telescope Science Institute, 3700 San Martin Road, Baltimore, MD, 21218, [4]Lowell Observatory, 1400 West Mars Hill Road, Flagstaff, AZ, 86001.

**Our discovery of two new satellites of Pluto[1], designated S/2005 P 1 and S/2005 P 2 (henceforth, "P1 and "P2"), combined with the constraints on the absence of more distant satellites of Pluto[2], reveal that Pluto and its moons comprise an unusual, highly compact, quadruple system. The two newly discovered satellites of Pluto have masses that are small compared to both Pluto and Charon, i.e., between $5 \times 10^{-4}$ and $1 \times 10^{-5}$ of Pluto's mass ($M_{pl}$), and between $5 \times 10^{-3}$ and $1 \times 10^{-4}$ of Charon's mass. These facts naturally raise the question of how this puzzling satellite system came to be. Here we show that P1 and P2's proximity to Pluto and Charon, along with their apparent locations in high-order mean-motion resonances, likely result from their being constructed from Plutonian collisional ejecta. We argue that variable optical depth dust-ice rings form sporadically in the Pluto system, and that rich satellite systems may be found—perhaps frequently—around other large Kuiper Belt objects.**

Here we report on the implications of the discoveries of two new satellites of Pluto[1]. The orbits of P1 and P2 reveal that Pluto's satellite system is both largely empty and highly compact (see Figure 1). All three of Pluto's known satellites orbit in the inner ~3% of Pluto's satellite prograde orbit stability radius[3], which extends outward to 2.2x10^6 km from Pluto. Outside of the three satellite orbits, the system appears to be devoid of other bodies[2].

We calculated the characteristic tidal e-folding spin-down time[4] $T^{spindow}$ for these bodies, to evaluate whether they should be expected to have rotational periods similar to their weeks-long orbital periods.  We assumed standard values for the mass and radius of Pluto[5], and minimal masses of P1 and P2. For P1 we found $T_1^{spindown}$=6.1x10^9 $(Q_1/k_2)$ yr, where $Q_1$ is the dissipation factor for P1 and $k_2$ is the assumed second degree potential Love number. For P2 we found $T_2^{spindown}$=5.3x10^8 $(Q_2/k_2)$ yr, where $Q_2$ is the dissipation factor for P2.  Because $Q_1/k_2$>20 for any non-pathological case, and in fact is most likely of order 10^4, it is clear that the characteristic spin-down times for both P1 and P2 are expected to significantly exceed the 4.5 Gyr age of the solar system.  As a result, P1 and P2 are not expected to be in synchronous rotation with Pluto unless they previously orbited much closer to Pluto, where the spin-down time is decreased by orders of magnitude.  If P1 or P2 are someday discovered to be spin-orbit synchronized, it would therefore suggest that these satellites formerly spent some considerable time closer to Pluto and then subsequently migrated outward as Charon and Pluto exchanged orbital and spin angular momentum to reach their current tidal equilibrium state. We return to this point later when discussing the origin of the system.

First, however, we develop some collisional considerations. Studies of the collisional environment of the present-day KB acting on Pluto-Charon and KB bodies of smaller sizes revealed[6,7] that the critical size boundary for catastrophic breakup over the past 4 Gyr occurs at diameters of ~4 km. P1 and P2 are large compared to this critical size scale for catastrophic disruption in the Kuiper Belt. P1 and P2 are thus likely to be ancient bodies originally formed during the same era as Pluto and Charon, and are unlikely to have been subsequently disrupted and re-accreted in the past few Gyr.

Collisional studies have also revealed[6] that in the current day KB, the cumulative fraction of the surface cratered by all 8 m diameter and larger KBO impactors ranges from ~7% to ~32% for bodies on orbits approximately like Pluto's. This does not include the additional surface area covered by ejecta blankets, which would increase this by a factor of 2 to 4, nor does it take into account higher cratering rates in the ancient KB, prior to its mass depletion. Even for these conservative assumptions, we can predict that the surfaces of P1 and P2 will be significantly cratered.

Characteristic collisional velocities for KB impactors onto these satellites are in the range 1-2 km s$^{-1}$. It has been demonstrated[6] that objects of the size class of P1 and P2 in the KB have likely lost some 10% to 20% of their mass to impact erosion. We therefore conclude that the present day sizes and mass of P1 and P2 are not very different from their sizes and masses at the time of their formation.

The characteristic ejecta velocity resulting from collisions onto these satellites should be of order 1% to 10% of the speed of Kuiper Belt debris impactors, or 10-100 m s$^{-1}$. At these speeds, collisional ejecta fragments will escape the satellites themselves but generally remain trapped in orbit about Pluto. This is in contrast to the situation obtaining at Charon (with its ~500 m s$^{-1}$ escape velocity): most collisional ejecta falls back onto Charon's surface, and does not reach orbit about Pluto. As such, the bombardment of P1 and P2 by small Kuiper Belt debris almost certainly generates faint, dusty ice particle rings around Pluto, with time-variable optical depth.

A crude estimate of the crude optical depth of these rings can be derived by assuming that 10% of the mass of these satellites may have been eroded from them over time. If we then adopt the conservative assumptions of: (i) minimum satellite masses for P1 and P2, (ii) a mean lifetime for ejected particles of $10^5$ years (i.e., an order of magnitude shorter than their lifetime for the erosion and sublimation of KB dust particles[8]), (iii) that only $10^{-4}$ of the debris is in micron-sized particles, (iv) that the ring particles have 1 g/cm$^3$ density, and (v) a characteristic width spanning the entire separation between P1 and P2, we derive a characteristic ring optical depth estimate of $\tau=5\times10^{-6}$. This is comparable to the optical depth of Jupiter's tenuous ring system. This estimate is only very approximate, but from it we conclude that Pluto can transiently possess dust rings as a result of the stochastic bombardment of P1 and P2 by small Kuiper Belt debris.

Now consider how P1 and P2 may have formed. Pluto's satellite Charon is half of Pluto's diameter, and has a specific angular momentum so high that there is broad agreement that the pair was generated via a giant collision with an ancient impactor[9,10,11,12]. But what is the origin of P1 and P2, two remarkably smaller satellites exterior to Charon?

P1 and P2's proximity to Pluto and to Charon, along with their apparent locations in high-order mean-motion resonances in the plane of Charon's orbit[1], present challenges to any assumed capture origin, but naturally suggest a formation in association with the giant impact origin of Charon. We therefore suggest that P1 and P2 are, like Charon, likely to be constructed of material ejected from Pluto and/or the Charon progenitor.

This hypothesis is supported by the circular or near-circular orbits of P1 and P2. We elaborate on this case by estimating the characteristic e-folding eccentricity decay time[4]:

$$\tau = e/[de/dt] = 2Q_s/[21\mu n(R_s/a_s)^5 k_2].$$

Here $Q_s$ is the dissipation coefficient for the satellite, $\mu$ is the satellite mass ratio relative to its primary (Pluto in this case), n is the orbital mean motion of the satellite, $R_s$ is the satellite's radius, $a_s$ is the satellite's orbital semi-major axis, and $k_2$ is the second degree potential Love number of the satellite.

Adopting the satellite orbits we reported elsewhere[1], $Q_s$=100 (considered typical of icy satellites), $k_{2s}$=0.055 (appropriate for rigid ice[4]), densities of 2 g cm$^{-3}$ (i.e., similar to Pluto and Charon), and assuming that P1 and P2 have their maximum permissible radii[1], we find tidal circularization time scales of 65 Gyr and 500 Gyr, respectively. Thus, it is seen that near their current orbits, or farther out, the eccentricity decay times for P1 and P2 are far too long to damp from high eccentricity capture values to circular orbits in the age of the solar system unless either the satellite Q's are <1 (strengthless rubble piles) and/or unless gas drag assisted any putative eccentricity decay. In contrast, Charon's orbital eccentricity decay time scale is short, $\sim 3 \times 10^6$ yr, and the tidal decay time scale for eccentricity near Pluto's Roche lobe is only of order $10^5$ years. Occam's razor thus suggests that the circular orbits of P1 and P2 imply that (i) they most likely formed much closer to Pluto, rather than farther out or by capture from heliocentric orbit, and (ii) they subsequently evolved outward to their present-day positions during the tidal evolution of Charon to its current orbit.

This said, we caution that their very small masses relative to Charon beg the question of why so little material would have escaped accumulation into orbiting bodies other than Charon, and therefore, why are there not more small satellites of Pluto. Perhaps other satellites did form, but eventually became dynamically destabilized, resulting in accumulation onto Charon or Pluto; or perhaps there are other, still fainter, satellites that escaped detection below the new HST observation[2] threshold near V=26.2.

Finally, it has been estimated that 20%, or more, of the known KBOs have satellites[13]. This suggests that there must be tens of thousands of KBOs with satellites. Given this, and our discovery of P1 and P2 orbiting Pluto, we consider it likely that many more KBO satellite systems will be revealed to be multiples when examined more closely. Figure 1 further motivates this point, showing the ample space available for minor satellites in the known KBO systems.

We suggest that a natural place to expect multiple satellite systems (and associated rings) would be around those KBOs which possess tightly bound, large satellites reminiscent of binary formation events like the Pluto-Charon pair. Such objects include 1997 CQ29, 1998 SM165, 1999 TC36, 2003 UB313, and 2003 EL61. It will also be useful to search for more distant, irregular satellites orbiting KBOs to determine whether less compact, e.g. capture-related architectures also exist among KBOs with satellite systems.

**Figure 1. The architecture of the Pluto system is compared here to other KBOs with known satellites and to the Earth-Moon system.** The orbital distances and sizes of all three satellites in the Pluto system are shown here in comparison to other relatively well characterized KBO-satellite systems, and the Earth-Lunar pair. P1 and P2 orbit relatively close to Pluto at distances of 64,700±850 km and 49,400±600 km, respectively[1]. Photometry of these two bodies[1], indicates that their visual magnitudes were V=22.93±0.12 and 23.38±0.17, respectively, in mid-May 2005. For an assumed (i.e., comet-like) lower limit albedo of 0.04 (as shown), one derives upper limit diameters of 167±10 km for P1 and 137±11 km for P2. If their albedos are as high as 0.35 (i.e., like Charon[5], a reasonable upper limit), then their diameters are only ~61±4 km and ~46±4 km, respectively. Pluto apparently has no undiscovered satellites farther out in the system down to objects 40 times fainter than P1 or P2. For this figure, all satellites are assumed to lie at their discovery distance if a formal semi-major axis has not yet been established. The left-hand panel shows satellite sizes on an absolute scale, with orbital distances normalized to the orbital stability zone within which the primary body can retain satellites over long time scales. Masses were computed from sizes assuming a density of $\rho=2$ g cm$^{-3}$, like Pluto and Charon[5]. The right-hand panel shows the systems with satellite

sizes normalized to the radius of the primary in each system (e.g., Charon appears larger than the Moon), and their orbital distances in units of the primary's radius; object sizes were computed assuming 4% albedos.

**Author Information:** Reprints and permissions information is available at npg.nature.com/reprintsandpermissions. The authors declare no competing financial interests. Correspondence and requests for materials should be addressed to S.A.S (email: astern@swri.edu).

**Acknowledgements.** We thank Bill McKinnon and Robin Canup for reading and commenting on this manuscript.

**Figure 1**

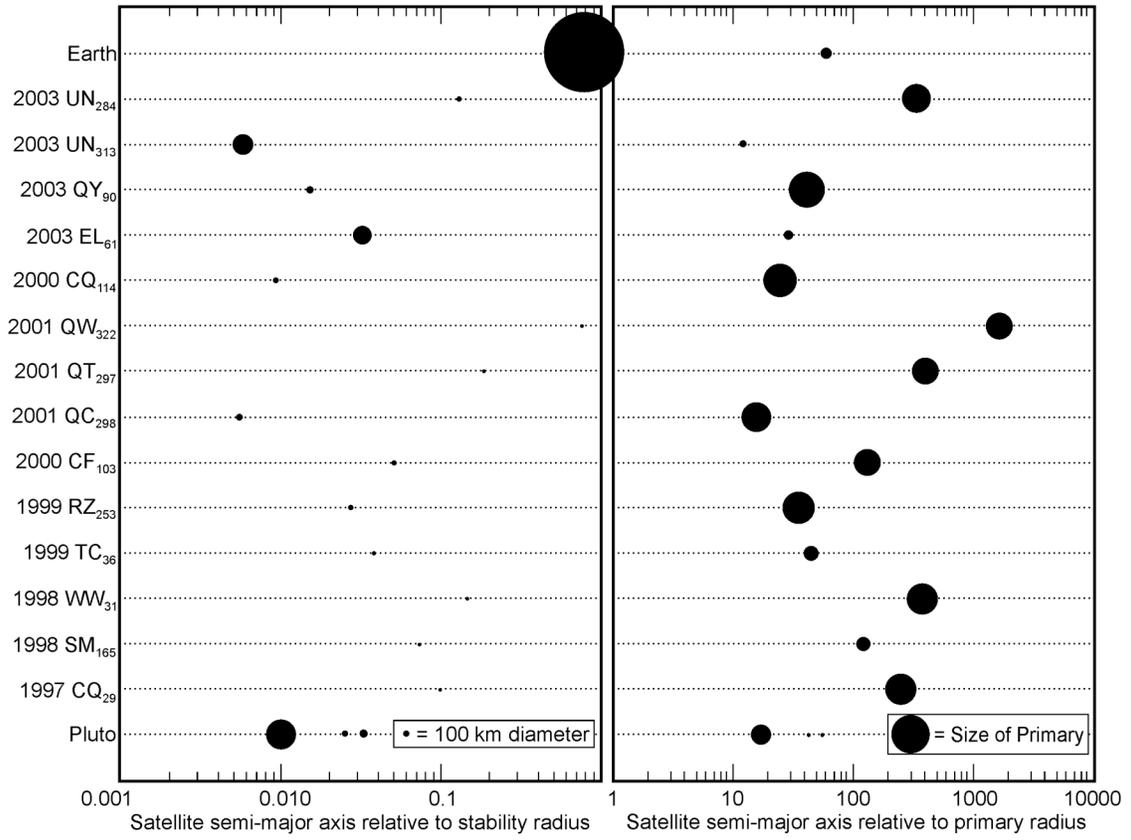